\documentclass[acmsmall]{acmart}

\usepackage{xcolor}
\usepackage{multirow}

\AtBeginDocument{%
  \providecommand\BibTeX{{%
    \normalfont B\kern-0.5em{\scshape i\kern-0.25em b}\kern-0.8em\TeX}}}

\copyrightyear{2022}
\acmYear{2022}
\setcopyright{rightsretained}
\acmJournal{PACMHCI}
\acmYear{2022} \acmVolume{6} \acmNumber{CSCW2} \acmArticle{456} \acmMonth{11} \acmPrice{} \acmDOI{10.1145/3555557}

\usepackage{etoolbox}
\makeatletter
\patchcmd{\maketitle}{\@copyrightpermission}{
   \begin{minipage}{0.2\columnwidth}
     \href{https://creativecommons.org/licenses/by/4.0/}{\includegraphics[width=0.90\textwidth]{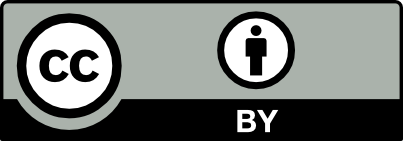}}
   \end{minipage}\hfill
   \begin{minipage}{0.8\columnwidth}
     \href{https://creativecommons.org/licenses/by/4.0/}{This work is licensed under a Creative Commons Attribution International 4.0 License.}
   \end{minipage}

   \vspace{5pt}
}{}{}

\makeatother

\begin{document}

\title[Revisiting Piggyback Prototyping]{Revisiting Piggyback Prototyping: Examining Benefits and Tradeoffs in Extending Existing Social Computing Systems}

\author{Daniel A. Epstein}
\email{epstein@ics.uci.edu}
\affiliation{\institution{University of California, Irvine}\country{USA}}

\author{Fannie Liu}
\affiliation{\institution{Snap Inc.}\country{USA}}
\email{fannie@snap.com}

\author{Andrés Monroy-Hernández}
\affiliation{\institution{Snap Inc. \& Princeton University}\country{USA}}
\email{amh@snap.com}

\author{Dennis Wang}
\email{dennisw7@uci.edu}
\affiliation{\institution{University of California, Irvine}\country{USA}}

\renewcommand{\shortauthors}{Daniel A. Epstein et al.}

\received{January 2022}
\received[revised]{April 2022}
\received[accepted]{May 2022}

\begin{abstract}
  The CSCW community has a history of designing, implementing, and evaluating novel social interactions in technology, but the process requires significant technical effort for uncertain value. We discuss the opportunities and applications of ``piggyback prototyping’’~\cite{grevet2015piggyback}, building and evaluating new ideas for social computing on top of existing ones, expanding on its potential to contribute design recommendations. Drawing on about 50 papers which use the method, we critically examine the intellectual and technical benefits it provides, such as ecological validity and leveraging well-tested features, as well as research-product and ethical tensions it imposes, such as limits to customization and violation of participant privacy. We discuss considerations for future researchers deciding whether to use piggyback prototyping and point to new research agendas which can reduce the burden of implementing the method.
\end{abstract}



\maketitle

\section{Introduction}

The CSCW research community has a long history of designing, implementing, and evaluating novel social communication technologies to explore how to address the limitations of existing systems or envision alternative futures. Through this work, we as a community aim to address classic CSCW problems like connecting people with similar interests and goals, helping them have conversations about important topics, minimizing harmful content, and more. As outlined in the joint UIST+CSCW panel at the 2020 conference~\cite{bernstein2020uist+}, the decades of systems research on collaborative and social computing have led to a richer understanding of how to design these systems and important improvements in the social tools we use daily.

It is well known that there are significant barriers to designing, implementing, and evaluating new systems and design ideas \textcolor{black}{to answer research questions about} social communication. For example, Bernstein et al. remark on how challenging it is to build up a social computing system, and how research incentives are often at odds with commercial incentives to do so~\cite{bernstein2011trouble}. Community events often bring together researchers to discuss these issues and develop paths forward, like the Social Computing Systems Camp~\cite{zhangsummercamp}, the CSCW 2021 panel discussion on the research community's role in reforming or remaking social technologies~\cite{thebault2021we}, and the yearly CrowdCamp workshop at HCOMP~\cite{andre2012crowdcamp}.

To address some of these concerns, Grevet \& Gilbert propose \textit{piggyback prototyping}, or building new ideas for social computing systems on top of existing systems, as one technique to more rapidly support evaluating design ideas prior to developing an entirely new social computing system~\cite{grevet2015piggyback}. In the years since this proposal, social computing platforms have increasingly incorporated resources and APIs which allow for the customization and modification needed to piggyback prototyping. For example, programmable and configurable bots are central to chat and work platforms like Slack, Discord, and WeChat~\cite{lin2016developers, lebeuf2017software}, online communities like Reddit and Twitch~\cite{jhaver2019human, seering2018social}, and peer production communities like Wikipedia~\cite{geiger2014bots}. Social computing systems also often allow people to share content they generate in other apps and tools, such as through share sheets provided by phone operating systems for sharing to social media or chat apps (e.g., Twitter, Instagram, iMessage, Google Chat) and more sophisticated APIs like Snap's Creative Kit and Lens Studio~\cite{snapcreativekit, snaplensstudio}. \textcolor{black}{Researchers often leverage these resources when implementing piggyback prototypes to answer their research questions, such as creating research apps that can post to participant's social media accounts to evaluate new kinds of content~\cite{munson2012exploring, suh2014babysteps, epstein2020yarn} or creating bots that can post messages to feeds where participants interact with one another to evaluate new content summarization or organization strategies~\cite{zhang2018making, salehi2018hive, seering2020takes}}.

With the benefit of hindsight, we revisit and expand on Grevet \& Gilbert's work by critically examining and discussing the benefits and tradeoffs of research which uses the method of designing, developing, and evaluating social communication experiences that piggyback off of existing systems in comparison to building them from scratch. We summarize and draw on observations from researchers who developed about 50 prior social computing systems \textcolor{black}{to answer their research questions}, using three case studies where we are more familiar with their design and evaluation to further describe how the method can support different research goals. While Grevet \& Gilbert suggest that piggyback prototyping can help assess the critical mass of interest prior to spending resources developing a new social computing platform, we surface that \textcolor{black}{the research community} has additionally used piggyback prototyping to \textcolor{black}{answer research questions about the design of} social computing systems. 
For example, piggyback prototypes have contributed to different onboarding and social support interventions for introducing new members to the norms of established social communities like Wikis and Q\&A sites ~\cite{narayan2017wikipedia, lu2018streamwiki, ciampaglia2015moodbar, ford2018we}. Research systems have also piggybacked to explore how designs can support sharing new kinds of information in feed-based conversations, like health data ~\cite{munson2012exploring, epstein2020yarn, munson2015effects} and biosignals~\cite{liu2017supporting, curmi2013heartlink, liu2018reactionbot}.


Drawing on the research literature's frequent use of piggyback prototyping to create recommendations for the design of social computing systems, we describe and discuss the benefits of the method and the tensions faced compared to developing social computing systems from scratch. We identify benefits which are \textit{intellectual}, such as improved ecological validity and the ability to study more complex social relationships. We also address \textit{technical} benefits, such as the ability to leverage existing well-tested features. At the same time, we characterize \textit{research-product tensions} such as the increased expectation of polish of all aspects of the system and limits to customization, as well as \textit{ethical tensions} around access to research data outside of the research team, violation of participant privacy, and exclusionary practices reinforced by piggybacking on platforms. In summarizing these benefits and tradeoffs, we advocate for more CSCW systems research to consider piggybacking as a method and for more CSCW systems to enable being piggybacked on by later researchers.

By critically examining piggyback prototyping systems, we contribute:
\begin{itemize}
    \item A review of design challenges for social computing systems where piggyback prototyping has been used to help address, extending Grevet \& Gilbert's focus on evaluating interest level~\cite{grevet2015piggyback}. We draw on about 50 past systems which have leveraged the method, characterizing five design challenges: creation and personalization of social content, ordering and filtering of content, establishing affinity groups, onboarding new members, and content moderation.
    \item Descriptions of core benefits and tensions of piggyback prototyping. We characterize four intellectual and technical benefits to evaluating design ideas for social computing systems within existing systems: in-situ evaluation, asymmetric participation and audience access, leveraging built-in mechanics, and simplified and organized participant data collection. We also articulate two main tensions around using the method: tensions between the goals of the research and the goals of the product being piggybacked off of, and ethical tensions around using existing platforms for research. In reflecting on three research projects, SnapPI, Everyday Biosignals, and Mindful Garden, we consider how piggyback prototyping improved aspects of these works while also introducing design and evaluation challenges.
    \item Recommendations for future researchers and platform creators interested in using or supporting piggyback prototyping. We describe three considerations for researchers deciding whether to use the method: whether platforms allow for the desired customization, whether people are already using the platform to communicate, and whether piggybacking introduces privacy or ethical concerns. We advocate for piggyback prototyping by building on prior arguments around the difficulties of CSCW systems research, and point to new research agendas, such as developing customizable authoring tools for existing social platforms.
\end{itemize}

\section{Redefining Piggyback Prototyping Social Computing Experiences}

Grevet \& Gilbert define \textit{piggyback prototyping} as a ``social computing system prototyping technique that utilizes existing social platforms to evaluate novel social interactions for large-scale systems''~\cite{grevet2015piggyback}. In describing stages of piggyback prototyping, Grevet \& Gilbert focus on using the method to evaluate whether an idea for a new community is likely to attract a critical mass of interested people prior to investing development effort. We broaden this definition to include \textit{prototyping a novel social interaction utilizing an existing platform, where the intent is to address a common communication problem, barrier, or challenge or introduces a new communication capability, opportunity, or vision}. In addition to measuring interest in an idea prior to scaling up development, our definition includes prototyping with the goal of creating recommendations for improving social computing systems which either already exist or could exist in the future. The recommendations could aim to improve a specific system or platform (e.g., Facebook, SMS) or systems that have shared characteristics (e.g., systems with ephemeral messaging, systems where people have large audiences).

For this paper, we focus on the potential for piggyback prototyping to support research questions around improving a person's \textit{social or communication practices} in a system, as understanding and supporting socialization and communication is a fundamental research goal of the CSCW community. This distinction leaves many social computing systems that have built on existing systems outside of the scope of our discussion. For example, research tools have often piggybacked off of social computing systems to promote other goals, such as embedding microtasks into social feeds for productivity~\cite{hahn2019casual} or temporarily blocking or altering social websites to encourage a person to do other activities~\cite{agapie2016staying, kovacs2018rotating, kim2017technology, lyngs2020just}. The interventions of these systems may have a side-effect of potentially reducing how frequently people use that system to communicate, but modifying social interaction was not the primary intent. Similarly, research systems in Crowdsourcing have piggybacked off of existing systems (e.g., Amazon Mechanical Turk, Upwork) to evaluate different workflows or structures~\cite{bernstein2010soylent, kittur2011crowdforge, retelny2014expert}, but these systems, by in large, have not modified how workers communicate with one another. These examples highlight that piggyback prototyping can support a wide range of research goals beyond those we discuss. 

CSCW research also frequently examines the impact of changes that social platform designers make on people's experiences with the platform. For example, Dasgupta \& Hill used a policy change in the Scratch community to causally evaluate the impact of enabling cloud variables on learners' sharing behavior~\cite{dasgupta2018wide}. Evaluations are occasionally planned experiments around design features organized by the platform, such as Facebook's social contagion study~\cite{kramer2014experimental}. The benefits and challenges of evaluating novel social computing systems has overlap with experimentation that the systems' creators or maintainers may perform, such as A/B testing. However, most of the implications we discuss in this work relate to designs created and evaluated by researchers outside the platforms, and who may have limited ability to modify the platform through available APIs or have partnerships with platform development teams.

Many systems research contributions, including social computing systems, do not necessitate eliciting formal feedback from the people for whom the technology is designed. Proof-by-existence, articulating a vision, or improving technical performance may all be more appropriate evaluations for many research contributions~\cite{olsen2007evaluating, greenberg2008usability}. Many of the benefits and challenges we articulate for piggyback prototyping are specific to evaluating the prototype in-situ with the population it was designed to support, such as through a deployment or a laboratory study. We also highlight that towards many research questions, understanding people's current social communication practices on an existing platform may be sufficient or more valuable than designing a piggybacked system that tests a different communication practice. We do not advocate for piggyback prototyping as a replacement for these sorts of contributions.

\section{Research Approach}

Building an understanding of the design challenges piggyback prototyping has been used to address, and the benefits and tradeoffs of the method, requires identifying a set of literature that contributes piggyback prototyped systems. To the best of our knowledge, there are no formal lists of literature or common keywords used in the literature to describe piggyback prototyping, removing the option of conducting a systematic literature review. At the time of writing, searching for ``piggyback prototyping'' and ``piggyback prototype'' in the ACM digital library resulted in 3 unique articles. Although we have drawn significant inspiration from Grevet \& Gilbert's characterization~\cite{grevet2015piggyback}, relatively few systems that have since built on existing social computing systems have cited the article. 
Similarly, searching for the names of commercial systems like Snapchat or Reddit would result in significantly more articles that do not meet the inclusion criteria than those that do, such as the study of people's use of the platforms.

To identify the literature that used the method, we, therefore, began with a list of about 20 articles we were aware of from our own experience conducting research in the space. Over the course of a year, we added to this list by (1) reading the literature that cite and are cited by those 20 articles (and looking at subsequent references), (2) examining the publication histories of the scholars who wrote the original articles, (3) perusing the recent and upcoming proceedings of prominent venues like CSCW, CHI, and UIST, and (4) asking our social networks. Through this process, we also refined our definition of piggyback prototyping of social computing experiences. In total, we identified about 50 articles which contributed piggyback prototyped systems. We identified piggybacking systems from as early as 2004, with most having been published in the last decade.

We do not claim that this list is complete, but rather that the benefits and challenges discussed by those articles reflect that of the method applied generally. \textcolor{black}{Within and outside of CSCW, there is inevitably a great deal of research that piggybacks off of commercial platforms like Google Docs or Instant Messaging clients. At scale, it can be difficult to differentiate research systems which aim to improve social and communication practices from those which aim to support other goals. While conducting a more thorough review is outside the scope of this research, critically examining how a particular platform or class of platforms (e.g., Google Docs or collaboration tools) have been extended to provide different social experiences would generate useful guidelines for toolkits looking to reduce the threshold to creating these sorts of systems.}

To understand researchers' perceptions of piggyback prototyping, we analyzed the system description, method, discussion, and limitations sections of the articles we identified. We specifically looked for passages which described: (1) what intended benefits influenced the choice to develop a piggyback prototype rather than design a system from scratch, and (2) what limitations, downsides, or concerns they encountered with the method, if any. In total, we identified about 100 passages which articulated benefits or limitations, ranging from a few sentences to multiple paragraphs.

\textcolor{black}{Following a thematic analysis~\cite{braun2006using} approach, the first author began by revisiting the passages and open coding each. The 15 codes included ``\textit{benefits of being a `real world' system}'', ``\textit{needing to fit with a sharer's platform preferences}'', and ``\textit{consideration of terms of service}''. Through multiple discussions as a research team, we then consolidated our codes and grouped these passages, naming and categorizing the benefits and tensions identified in prior research}. Table ~\ref{tab:examples} contains the full list of articles we identified, summarizing the existing systems that we commonly saw piggybacked on.

\begin{table}[ht]
\caption{Examples of existing social systems on which the research literature has piggybacked from our literature review.}
\scalebox{0.9}{%
\label{tab:examples}
\begin{tabular}{|p{0.35\linewidth}|p{0.65\linewidth}|}
\hline
\textbf{Existing system} & \textbf{Exemplar piggyback prototype(s)} \\ \hline
Facebook & Asynchronous Remote Communities (ARCs)~\cite{macleod2016asynchronous, prabhakar2017investigating, maestre2018defining}, CommitToSteps~\cite{munson2015effects}, FeedVis~\cite{eslami2015always, eslami2016first}, Food4Thought~\cite{epstein2016crumbs}, GoalPost~\cite{munson2012exploring}, HeartLink~\cite{curmi2013heartlink}, Regroup~\cite{amershi2012regroup}, Three Good Things~\cite{munson2010happier}, Yarn~\cite{epstein2020yarn} \\ \hline
SMS, Chat, and IM Apps (e.g., WhatsApp, WeChat, KakaoTalk) & Chorus~\cite{huang2016there}, DearBoard~\cite{griggio2021mediating}, Everyday Biosignals~\cite{liu2017supporting}, Mental Health Chatbot~\cite{lee2020designing}, ReactionBot~\cite{liu2018reactionbot}, TableChat~\cite{lukoff2018tablechat}, MyButler~\cite{cho2020share}, WeRun~\cite{li2020supporting}, Whatfutures~\cite{lambton2019whatfutures, lambton2021blending} \\ \hline
Wikis and Forums & ForumReader~\cite{viegas2006visualizing}, MoodBar~\cite{ciampaglia2015moodbar}, ORES~\cite{halfaker2020ores}, Snuggle~\cite{halfaker2014snuggle}, The Wikipedia Adventure~\cite{narayan2017wikipedia}, Wikipedia Teahouse~\cite{morgan2013tea} \\ \hline
Email & Bluemail~\cite{whittaker2011wasting}, Calendar.help~\cite{cranshaw2017calendar}, Squadbox~\cite{mahar2018squadbox}, Themail~\cite{dave2004flash} \\ \hline
Streaming Platforms (e.g., Twitch, YouTube Live, DouYu) & BabyBot~\cite{seering2020takes}, Snapstream~\cite{yang2020snapstream}, StreamSketch~\cite{lu2021streamsketch}, StreamWiki~\cite{lu2018streamwiki}, VisPoll~\cite{chung2021beyond} \\ \hline
Slack & Hive~\cite{salehi2018hive}, T-Cal~\cite{fu2018t}, Tilda~\cite{zhang2018making} \\ \hline
Twitter & @BabySteps~\cite{suh2014babysteps}, Airport Check-ins~\cite{grevet2015piggyback}, WeMeddle~\cite{gilbert2012predicting} \\ \hline
Snapchat & Mindful Garden~\cite{mindfulgarden}, Project IRL~\cite{dagan2022project}, SnapPI~\cite{snappi} \\ \hline
Reddit & CrossMod~\cite{chandrasekharan2019crossmod}, PolicyKit~\cite{zhang2020policykit} \\ \hline
Google Docs& Guided Chat~\cite{o2018suddenly}, WearWrite~\cite{nebeling2016wearwrite} \\ \hline
YouTube & OtherTube~\cite{bhuiyan2022othertube} \\ \hline
Stack Overflow & Help Room~\cite{ford2018we} \\ \hline
Figma & Winder~\cite{kim2021winder} \\ \hline
JCow & PhamilySpace~\cite{sandbulte2021working} \\ \hline
\end{tabular}
\vspace{-13pt}
}
\end{table}

\section{Design Challenges Piggyback Prototyping has Been Used to Address}

To show how piggyback prototyping has utility beyond assessing the level of interest in a social computing system, we characterize the design challenges and visions that prior research has leveraged the method to produce. To inform scholars considering the method, we further describe its benefits and challenges relative to developing social computing systems from scratch.

We identified five challenges in the design of social computing systems where researchers have leveraged piggyback prototyping to develop recommendations. These challenges do not necessarily articulate all design challenges where researchers use piggyback prototyping, but characterize some common CSCW challenges where the method is useful. In outlining these challenges, we expand the method's applications beyond Grevet \& Gilbert's original intentions for it to evaluate interest in a social computing system~\cite{grevet2015piggyback}.

\subsection{Supporting Creation and Personalization of Social Content}

Researchers often extend social computing systems to enable the sharing of new or rarely shared content, evaluating people's perspectives of that content. For example, research systems have supported sharing sensed data to social platforms, including biosignals like heart rate~\cite{curmi2013heartlink, liu2017supporting}, visual expressions as sensed by a webcam~\cite{liu2018reactionbot}, and steps or other forms of physical activity~\cite{munson2012exploring, munson2015effects, epstein2020yarn}. While it may be possible for people to share this content with enough effort (e.g., by copying text or taking a screenshot of another app), systems often aim to lower the barrier to creating this content. As another example, Calendar.help sought to reduce the burden of creating meeting invitations that work for multiple people's schedules, offloading scheduling efforts over email to a bot~\cite{cranshaw2017calendar}. While scheduling meetings and sharing invitations is widely done without Calendar.help, the system aimed to lower the burden.

Researchers also often extend social computing systems to examine how designs can encourage people to share content which is already commonly shared. For example, @BabySteps prompted people to share about their newborn’s development on social networking sites to more readily engage people in platforms they are already using~\cite{suh2014babysteps}. Three Good Things leveraged a positive psychology intervention, encouraging sharing positive behaviors on Facebook
with the hope of creating stronger bonds among close ties~\cite{munson2010happier}.

Other systems have supported personalizing or customizing how content appears or is delivered. DearBoard enabled people to personalize keyboards in messaging apps to promote connectedness~\cite{griggio2021mediating}. MyButler enabled scheduling timing of messages in KakaoTalk, allowing people to adjust when they shared and received messages~\cite{cho2020share}.

\subsection{Ordering, Filtering, and Summarizing Content}

CSCW systems often aim to improve how social content is ordered or summarized. Systems like Tilda~\cite{zhang2018making} and T-Cal~\cite{fu2018t} enabled tagging and summarizing group chats, evaluating their effectiveness by studying past and present group conversations on Slack. Similarly, systems like ForumReader and Themail aimed to improve forum organization and email exchanges to allow people to better understand the content and social dynamics~\cite{viegas2006visualizing, dave2004flash}.

Other systems have modified algorithms underlying how social content is ordered or filtered, such as Facebook's social contagion study~\cite{kramer2014experimental}. FeedVis similarly surfaced what content is shown and hidden in NewsFeed algorithms, enabling participants to inspect the algorithms' influence on the posts they received~\cite{eslami2015always, eslami2016first}.

\subsection{Establishing or Evaluating Affinity Groups}

Other research has leveraged existing social platforms to evaluate strategies for bringing people together around a shared topic or interest. For example, the Asynchronous Remote Community (ARC) method convenes groups on online platforms such as Facebook to participate in design workshops, contributing recommendations for systems looking to foster discussion on personal or sensitive topics like pregnancy and HIV~\cite{macleod2016asynchronous, prabhakar2017investigating, maestre2018defining}. Lambton-Howard similarly used WhatsApp to bring together a peer support community around health after learning how central WhatsApp was to other forms of communication~\cite{lambton2021blending}. Other systems have piggybacked to evaluate strategies for promoting competition on a shared topic, such as Food4Thought incorporating daily food challenges into a Facebook group to promote food mindfulness~\cite{epstein2016crumbs}.

Systems have also extended platforms for small group communication to steer conversations towards particular topics. For example, TableChat and WeRun both extended WeChat to encourage family conversations around healthy eating~\cite{lukoff2018tablechat} and physical activity~\cite{li2020supporting} through sharing data on those topics (e.g., food journal photos, step logs). In addition to contributing design strategies for having conversations, evaluation of these piggyback prototypes also helps contribute understanding how people use digital systems to communicate around these topics.

Researchers have also built on existing social platforms to explore approaches to changing the formation of the groups people chat and share in. For example, the Slackbot in Hive reconfigured groups of designers to better support collective design~\cite{salehi2018hive}. Systems such as ReGroup and WeMeddle supported people in creating social circles from among their existing friends on Facebook and Twitter~\cite{amershi2012regroup, gilbert2012predicting}.

\subsection{Onboarding New Members}

CSCW research, particularly in the space of peer production, has often prototyped approaches to address the challenge of introducing and supporting more diverse people to online communities. For example, Teahouse, embedded into Wikipedia as an alternative to the established help system, introduced a mentorship system to help new editors receive more support~\cite{morgan2013tea}. The Wikipedia Adventure gamified learning editing skills towards the same goal, recruiting participants from commenters on active Wikipedia pages~\cite{narayan2017wikipedia}. Other work has aimed to introduce skills, behaviors, or practices to new members, such as MoodBar looking to increase socialization~\cite{ciampaglia2015moodbar} and Snuggle normalizing critique practices~\cite{halfaker2014snuggle}. Outside of peer production, systems have aimed to help new users understand and follow community norms, such as Help Bar for formulating questions on Stack Overflow~\cite{ford2018we}.

\subsection{Content Moderation and Quality Control}

In large online communities that generate a lot of content, moderating content has emerged as an area where increased system support can help manage the workload. For example, tools like CrossMod and ORES have leveraged machine learning to initially evaluate content from existing online communities like Subreddits and Wikis~\cite{chandrasekharan2019crossmod, halfaker2020ores}. On a smaller scale, Squadbox explores asking a person's friends to evaluate email content they receive to avoid online harassment~\cite{mahar2018squadbox}. Evaluation techniques for these content moderation tools have varied, including recruiting participants to use the tool over a short duration~\cite{mahar2018squadbox}, mirroring posted content in a separate instance with the tool to get feedback from moderators~\cite{chandrasekharan2019crossmod}, and publicly releasing the tool and studying its use~\cite{halfaker2020ores}.

\section{Benefits of Piggyback Prototyping for Researchers}

Piggybacking off of existing social computing systems offers substantial benefits for researchers, including promoting more realistic evaluations, supporting varied study participation arrangements, and lowering the burdens of some aspects of development. We note that some of the benefits provide intellectual advantages like increased ecological validity, while others provide technical advantages like simplifying the development or deployment of a prototype.

\subsection{Intellectual benefits}

Piggyback prototyping enables answering research questions that would be more challenging or impossible to answer from deploying standalone systems, improving researchers' ability to understand how people use or perceive social tools in their everyday lives.

\subsubsection{In-situ evaluation}

Systems leveraging piggyback prototyping extend the overall benefits of field studies relative to other methods, such as helping avoid ecological gaps in understanding how people use technology in everyday life~\cite{siek2014field, thomas1989minimizing}. In social computing systems, people often have well-defined practices around what app(s) they use to communicate with others. Griggio et al. argue that when deploying a new communication tool, not requiring participants to relocate their conversations can allow for closer observation of the circumstances participants do and do not value the tool~\cite{griggio2021mediating}. Cho et al. similarly state that requiring participants to switch platforms would ``\textit{inevitably bring a new factor to the user's messaging experience}''~\cite{cho2020share}.

Evaluating in existing social contexts can also help participants more critically consider whether the extension helps address the communication challenge it was intended to address. For example, the evaluation of StreamWiki within an existing streaming platform (DouYu) pointed to a need for systems to better interact with other technologies people use for editing videos and communicating with their streaming audiences~\cite{lu2018streamwiki}. Similarly, participants using systems like GoalPost, Yarn, and CommitToSteps were faced with considering whether their audiences (in these cases, on Facebook) would be interested in that content~\cite{munson2012exploring, epstein2020yarn, munson2015effects}. It may be possible to derive parts of these insights from formative interviews or speculative studies, but deciding whether a proposed approach aligns with or improves existing communication practices deepens this understanding.

Integrating with existing platforms that participants are already using can help minimize the gap between who they communicate with during and outside the study. In deciding to piggyback off of certain systems, researchers frequently pointed to the market share of those systems or participants' level of familiarity as important factors~\cite{lukoff2018tablechat, cho2020share, liu2018reactionbot, lee2020designing, lambton2019whatfutures, lu2018streamwiki}. For example, TableChat chose to piggyback off of WeChat because an overwhelming majority of formative study participants already used WeChat for communicating with family members~\cite{lukoff2018tablechat}. Similarly, some iterations of the ARC method leveraged Facebook because people already formed groups on the platform for discussing the relevant health topics, leading to similar experiences in- and out-of study~\cite{macleod2016asynchronous}.

Building off of an existing system can also serve as a reminder for study participation, adherence, or commitment to behavior change. 
For example, Food4Thought integrated with a private Facebook group where study participants could communicate with one another~\cite{epstein2016crumbs}. Comments from other participants would often show up in their Facebook feed, reminding them to engage as they used the platform for other purposes. Three Good Things similarly hoped that participants' friends might remind them to engage with the intervention if they noticed that they were not posting on Facebook~\cite{munson2010happier}.



\subsubsection{Asymmetric participation and audience access}

Leveraging existing social computing systems enable studies where participants can be involved with different levels of engagement, and even engage with non-participants. 
This practice is often beneficial for systems that support creation of new types of content, as the person generating the content can then share it on the existing social platform. Studies that do not leverage existing social networking sites often instead require participants sign up for the study in pairs or groups, which results in limitations around group structure~\cite{wu2021receptivity, kim2020messaging}. Some studies also note the limitation of not capturing participants' typical communication practices, highlighting that participants wished they could use the standalone system to communicate with people who were not part of the study~\cite{liu2019animo, wu2021receptivity}. Asymmetric engagement also allows for snowball sampling, wherein an audience member can notice content shared by a study participant and decide to enroll in the study. 

For some system designs, the novel components being evaluated primarily impact only one person, allowing for others using the social system to be unaware of the study. For example, the Bluemail email client contributed new approaches for email searching and categorizing, requiring little need for engagement with the people with whom participants were sending or receiving emails~\cite{whittaker2011wasting}. Similarly, when FeedMe introduced a new technique for directed content sharing, it allowed recipients to receive shared content over emails they already used rather than sign up for the service themselves~\cite{bernstein2010enhancing}. Conversely, other systems have benefitted from reporting on the perspectives of multiple stakeholders. For example, in evaluating the Yarn system for structured storytelling on social media, Epstein et al. asked field study participants to disclose who they shared the content with, and sent those audience members a short survey on their experience receiving the content~\cite{epstein2020yarn}.

\subsection{Technical benefits}

As suggested by Grevet \& Gilbert piggyback prototyping avoids needing to reinvent features already present in existing social systems, such as social mechanics and participants' conversation logs. We further note benefits to the research, like participant familiarity with existing mechanics and simpler analysis.

\subsubsection{Leveraging built-in mechanics}
Compared to systems intended for individual use, systems with communal and collective mechanics often require significant development to ensure persistence of information across multiple people, features for socially engaging with one another (e.g., messages, likes, comments), methods for creating and customizing content, and more. 
Leveraging existing versions of these features can allow the development effort to focus on the interactive components most relevant to the research questions. Researchers often mentioned the benefit of a lower development and management burden for the research team, versus creating a social system from scratch~\cite{sandbulte2021working, grevet2015piggyback, macleod2016asynchronous}. \textcolor{black}{Importantly, this benefit can lead to faster iteration on design ideas and evaluation of them, allowing the research community at large to more rapidly design, evaluate, and share ideas for improving social computing systems~\cite{dow2009efficacy}.}

Leveraging mechanics from existing systems also promotes familiarity, both in overall interaction with a social computing system and specific features. 
For example, Tilda~\cite{zhang2018making} leveraged the social mechanisms of custom slash commands and emoji reactions on Slack, while Three Good Things ~\cite{munson2010happier} leveraged liking and commenting built into Facebook. PhamilySpace highlighted that building on an existing system ensured participant familiarity with the technology, allowing the evaluation to focus on social interactions rather than platform design choices~\cite{sandbulte2021working}. Novelty effects are also common in system deployment, where interest and engagement are high at the beginning of the study and taper over time. Minimizing new mechanics can possibly reduce this effect, as the design introduces less novelty overall.

Building off of existing systems also enables participants to continue using their typical communication strategies alongside the new strategy being evaluated. For example, in the UI design tool Winder, Kim et al. note that participants valued being able to use their typical chat alongside the new communication channel Winder introduced, leveraging the benefits of both~\cite{kim2021winder}.

\subsubsection{Simplified and Organized Participant Data Collection}
Particularly for public and semi-public social systems like some social media, piggybacked systems frequently touted the ease of collecting participant data. For example, in leveraging Twitter, Grevet \& Gilbert note the value of being able to log participant response or lack of response to their intervention via Twitter's APIs~\cite{grevet2015piggyback}. Other studies similarly describe the benefit of analyzing timestamped and linked Facebook messages from participants and audience members~\cite{curmi2013heartlink} or full chat logs from Twitch~\cite{seering2020takes}. Within the ARC Facebook groups, Macleod et al. comment that the sheer volume of participant data (comments, likes, timing of messages) provides rich insight into participant experiences, but can even make the analysis overwhelming without organization and structure~\cite{macleod2016asynchronous}.

\section{Tensions for Researchers when Piggyback Prototyping}

Though piggyback prototyping has significant benefits, utilizing the method also poses difficulties for researchers. Many challenges are around the tension between the nature of the piggybacked system as a product, as well as ethical questions resulting from leveraging an existing system.

\subsection{Research-product tensions}

Most of the systems that researchers piggyback off of are commercial applications, concerned with addressing the needs of their wide user base over the needs of researchers. Study participants using piggyback prototypes may extend their usability and performance expectations for commercial applications to the prototypes, increasing the need for them to be polished. Researchers also often wish to customize commercial tools in ways that are understandably not supported through APIs, limiting what piggyback prototypes they can create.

\subsubsection{Expectation of polish}

Because participants use piggybacked social computing systems to share content with non-participants, the expectations around system polish are often higher. For example, participants have expressed a need for many customization options to avoid repetitiveness~\cite{epstein2020yarn}, or critiqued the visual appearance of the content generated by the research system~\cite{munson2015effects}. While framing the research study goals and setting expectations for participants may allay some of these concerns, they inevitably impact participants' use and impressions of the system and the ability to evaluate the effectiveness of design approaches.

Leveraging systems where people are already communicating further requires care to ensure that the intervention does not interfere with participant's other communication. For example, in Squadbox, Mahar et al. argued that if anything were to go wrong (e.g., emails did not get delivered), it would negatively impact their participants~\cite{mahar2018squadbox}. They thus had to test their system extensively, because losses would not only impact their research but the well-being or productivity of their participants or others they communicate with. When designing and evaluating BabyBot, a chatbot for Twitch that learns from how others converse, Seering et al. had to anticipate what problematic phrases people might attempt to make the bot learn, and develop strategies to moderate and mitigate that risk to avoid subjecting audience members to inflammatory remarks~\cite{seering2018social}.

Piggyback prototypes often build on top of a single social platform to reduce the development burden. However, people often have idiosyncratic communication practices, defining rules for what app they use to communicate with whom and about what~\cite{nouwens2017whatsapp}. Participants using piggyback prototypes may find that the piggybacked platform does not align with the communication channel they prefer to use for that content~\cite{bernstein2010enhancing} or with other tools in their communication ecosystem~\cite{lu2018streamwiki}. 

\subsubsection{Limits to customization}

Many piggyback prototypes rely on clever workarounds or hacks to implement the social mechanisms they aim to evaluate. For example, some systems require disabling and replacing features of the systems they are prototyping on, such as disabling KakaoTalk's native notifications to allow MyButler to provide its own notifications instead~\cite{cho2020share}. Others require installing additional components developed by neither the researchers nor the piggybacked platform, such as a scheduling chatbot in TableChat~\cite{lukoff2018tablechat}. These workarounds also occasionally exploit potential privacy or security vulnerabilities with the piggybacked social systems or use their features in unintended ways. For example, DearBoard required taking a screenshot of a chat between two people in order to customize the keyboard for that interaction~\cite{griggio2021mediating}. Although all screenshots remained on the device, information about a person's conversation was still removed from the social interaction in order to implement the custom keyboard.

Technical barriers to customizing existing social tools also limit the kinds of interactions which can be supported. For example, in summarizing conversations on Slack with Tilda with a bot, Zhang et al. note that a more ideal interaction would introduce an overlay on top of or alongside existing conversation, but neither was possible~\cite{zhang2018making}. 
As a result, summaries took up significant space in chat streams, which they worried would be distracting for participants. Relatedly, in @BabySteps, Suh et al. point out that the character limit imposed by Twitter would limit prompts for participants to report on certain baby milestones~\cite{suh2014babysteps}. As a result, they had to adjust the wording of certain milestones, which could have impacted the reliability of the prompts.

Platforms often have policies which influence what kinds of interventions can be piggybacked. For example, when Ford et al. piggybacked Help Room on Stack Overflow, the platform's chat feature was only accessible to people who had sufficient reputation on the platform (e.g., 20 reputation points), interfering with the intervention's intent to help novices~\cite{ford2018we}. ReGroup notes specific evaluation choices required to stay within Facebook's Terms of Service~\cite{amershi2012regroup}. Other research systems point to technical limitations of APIs for piggybacking which may negatively impact participants' experience with the system, such as rate limits~\cite{eslami2015always, mahar2018squadbox, gilbert2012predicting} and unnecessary or distracting messages generated by the system being piggybacked on top of~\cite{cranshaw2017calendar, cho2020share}.

Platform limitations may also impact the choice of platform which can be piggybacked. In PolicyKit, available API endpoints limit what social platforms the tool could piggyback on top of~\cite{zhang2020policykit}. Zhang et al. remarked that Facebook Groups did not allow reverting or deleting messages, and changes were necessary to account for different audience response structures (e.g., emoji in Slack and Discord, up or downvotes on Reddit). 
Focusing on a single platform may limit generalizability of the design strategies proposed by a prototype, particularly if other platforms do not include similar capabilities or if the evaluation focuses on particular demographics who make up the bulk of the platform's users (e.g., young people, knowledge workers). For example, OtherTube piggybacked on YouTube via a desktop-only Chrome extension, which did not allow for understanding of the impact of its recommendation technique on mobile video browsing typical of young people~\cite{bhuiyan2022othertube}.

In addition to these customization barriers imposing limitations on platforms and interactions, they also limit the kinds of social interactions which are even possible to evaluate through piggybacking. Because we can only examine existing examples of piggyback prototyped systems, it is less evident what research questions have been abandoned or not explored due to a lack of platform support.




\subsection{Ethical tensions}

Repurposing systems that people use for everyday communication to instead be used for research introduces ethical questions. For researchers, this surfaces the question of how to balance participant privacy with the need to collect research data. At a broader level, it also surfaces a question around whether leveraging existing platforms needlessly subjects participants to the negative aspects of those platforms.

\subsubsection{Violating privacy expectations}

To answer research questions, the deployment of piggyback prototypes in research settings often necessitates observation of how, what, and with whom participants communicate using the prototype. Depending on the construction of the social system being piggybacked, access constraints can make it challenging to observe participant experiences. 
Grevet \& Gilbert highlight that the public nature of Twitter's feed proved helpful in understanding whether or how participants use the prototype to converse with others~\cite{grevet2015piggyback}. However, requiring participants to communicate on a public channel can also violate privacy. Suh et al. point out that while public participant actions can be beneficial for research observation, there is also risk when data being shared is sensitive, requiring informing participants of the risk or providing privacy controls~\cite{suh2014babysteps}.

Piggybacking off of systems that do not make conversations publicly available requires researchers to take alternate approaches to understand how participants use the system. Understanding participants' experiences can necessitate invading participant's social circles or conversations. In some studies, researchers chose to rely on participants to share their private experiences. WhatFutures designated a ``Future Guide'' whose goal was to collect what participants shared to the group~\cite{lambton2019whatfutures}. Other systems noted that a researcher joined the groups the systems created to observe and collect what participants were conversing about, such as private WhatsApp or Facebook groups~\cite{lambton2021blending, epstein2016crumbs}. In all examples we noted, this conversational data collection was disclosed and highlighted to participants ahead of time. Nonetheless, this research practice may come in tension with people's expectations around how conversations on that platform may be used.

Invading participant's private conversations may also influence what participants are willing to disclose. For example, researchers being in Facebook groups in the ARC method may have impacted participant's discussions on sensitive topics like HIV experiences~\cite{maestre2018defining}. Cho et al. similarly suggest that recording participant's KakaoTalk messages may have influenced the conversations they were willing to have with others~\cite{cho2020share}.

Piggyback prototyping introduces a risk of unintended data disclosures, both to the research team and to the platforms being piggybacked on top of. The design of piggyback prototypes may result in collecting data from people who did not explicitly consent to being a part of the study. For example, piggyback prototypes can produce posts to a public or semi-public social feed and analyze responses from audience members, such as in CommitToSteps~\cite{munson2015effects}. Although prototypes may clarify that the post was made as part of a research study and that engagement may be analyzed, Grevet \& Gilbert note the difficulties of obtaining explicit consent in the context of piggybacking and suggest that waivers of documentation may be sufficient~\cite{grevet2015piggyback}. However, many people do not believe that their public social media activity should be used without express consent~\cite{fiesler2018participant}.

Although participants may be comfortable sharing content with the research team, piggybacking on existing social media systems presents a risk of disclosing participant information to the social media platform~\cite{kosinski2015facebook}. Munson \& Consolvo expressed concern that their GoalPost app would disclose information about participants' exercise to Facebook, and therefore designed their piggybacking to include vague information in the Facebook posts themselves along with a link to a different site containing more details~\cite{munson2012exploring}.

\subsubsection{Reinforcing the negative aspects of social computing platforms}

Widely-used social computing platforms have faced criticism for promoting a range of negative behaviors, including inciting harassment, lowering self-esteem, and spreading disinformation~\cite{facebookfiles}. Requiring participants to engage with these platforms to use piggyback prototypes risks exposing them to this negative content during research studies. At a more meta-level, researchers building on and engaging with these platforms has the potential to provide endorsement of the platform's features and the creators' practices.

More directly, piggyback prototyping can potentially perpetuate exposure to negative content by promoting unethical practices. For example, researchers introducing or popularizing technical loopholes needed to conduct research work, such as publicizing a person's private activity, could make them more likely to be exploited by less benevolent parties. A social platform may further incorporate a participant's behavior in the piggyback prototype into later recommendations it provides (e.g., recommending advertisements influenced by their study activity, such as for HIV medication). Additionally, many people have opted out of commercial social platforms to avoid the content associated with those platforms. Piggybacking could therefore introduce a sample bias or a barrier to participation in research, limiting the generalizability of findings.

The CSCW community is continuing to discuss whether we should aim to iteratively improve existing social technologies to address these exclusionary practices or instead aim to develop new social technologies which do not encode them~\cite{thebault2021we}. While the community may agree on the overall goal (make social computing systems better for the people who use them), epistemological and philosophical perspectives may result in different perspectives on how to achieve that goal. Towards the vision of improving existing technologies, piggyback prototyping provides one potentially productive model for evaluating modifications to current social technologies which can improve on these negative experiences. However, it is important to acknowledge the limits of the method on reforming significant issues around these platforms.

\section{Considering Benefits and Drawbacks through Three Case Studies}

To further surface how the benefits and drawbacks of piggyback prototyping influence researchers' design choices, system development, and evaluation, we discuss three case studies of piggyback prototyping where we are familiar with the design and evaluation process. These case studies provide examples of how the method enables researchers to address design challenges that are otherwise difficult to address, but also introduce limitations impacting design or evaluation.

\subsection{SnapPI: Data-Driven Stickers for Ephemeral Social Media}

\begin{figure}[t]
    \footnotesize
      \centering
          \includegraphics[width=0.98\linewidth]{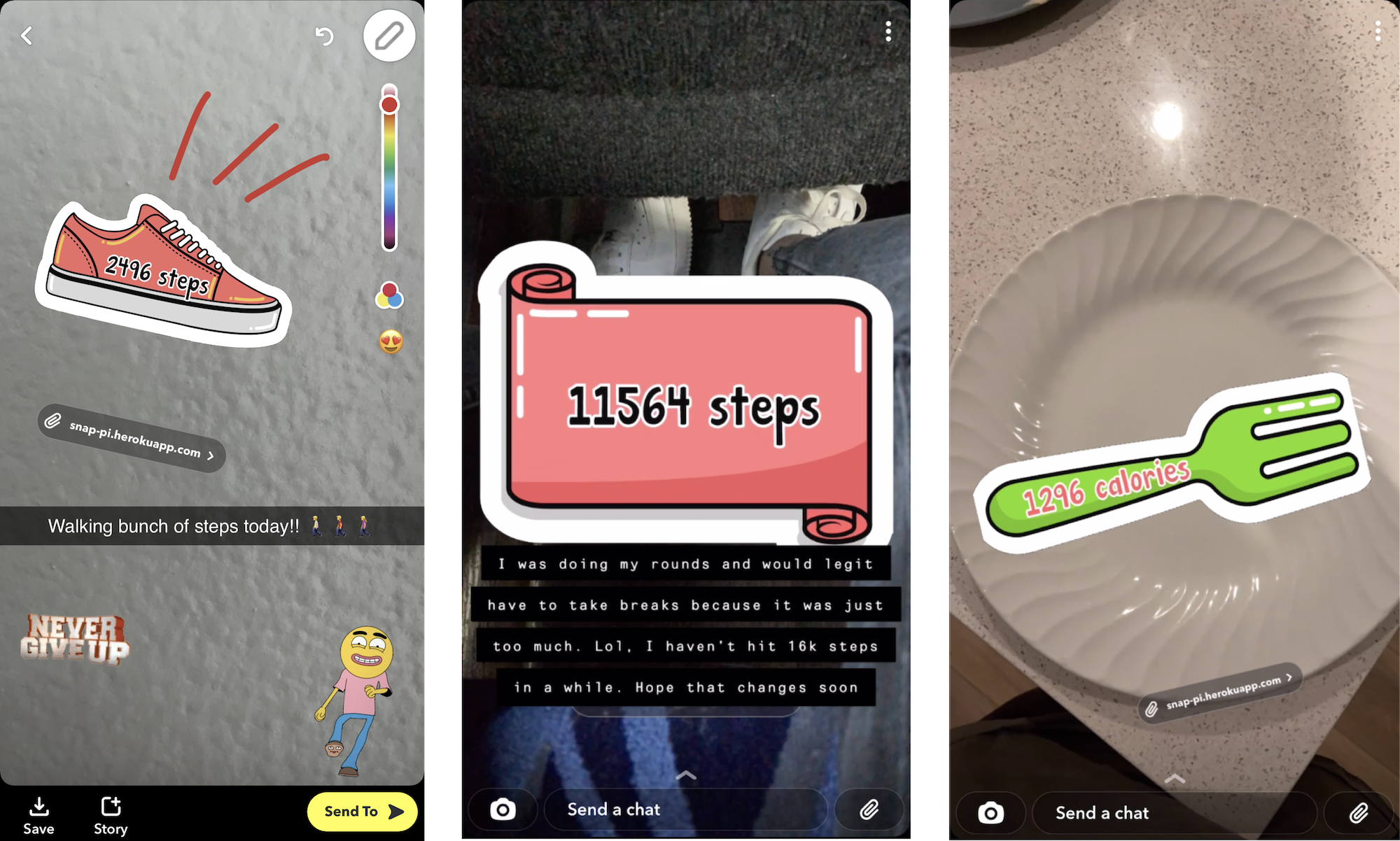} 
    \caption{SnapPI supported creating data-driven stickers for sharing on Snapchat. Piggybacking off of Snapchat allowed the researchers to leverage the platform's existing features, like annotation with other stickers and text, and supported understanding of how people wished to integrate stickers into everyday conversations.}
        \label{fig:snappi}
\end{figure}

Our first case study, SnapPI (Figure~\ref{fig:snappi}), supports creating stickers with personal data, such as physical activity and heart rate, for affixing to photos and videos shared on Snapchat~\cite{snappi}. Prior systems used piggyback prototyping to support sharing similar personal data to more archival social media (e.g., Facebook), finding that participants who used these systems were often concerned that the moments they collected with personal data were too trivial to share~\cite{munson2012exploring, munson2015effects, epstein2016crumbs, epstein2020yarn}. SnapPI was motivated by research suggesting that ephemeral platforms might be an effective space for balancing the benefits of sharing everyday data with concerns that the data does not represent significant enough accomplishments to warrant sharing on archival social media~\cite{epstein2020exploring}. Designing and evaluating SnapPI allowed for understanding when and how people would like to incorporate personal data into their everyday conversations with friends and family.

From a technical perspective, a key benefit of integrating into an existing platform enabled SnapPI to leverage many of the customization and sharing options that Snapchat provides. For example, participants playfully connected their data stickers with the photos that they shared. For example, a participant manipulated a ``fork'' sticker indicating how many calories they were eating to appear as though it was on top of the plate they had taken a photo of. These features enabled participants to transfer their communication style with how they use Snapchat's existing stickers to how they used SnapPI's data stickers.

Intellectually, the lack of friction when sharing with friends and family members deepened the researcher's ability to understand how people would like to use personal data to communicate. Participants appreciated how SnapPI stickers could naturally be integrated into the conversations they were already having on the platform. In particular, the researchers noted that the data sharing SnapPI supported led participants to have new conversations with their friends and family around health and wellbeing, as they were able to incorporate relevant stickers into the everyday conversations that they had on the platform.

In developing SnapPI, technical barriers influenced the design of the tool. Although multiple social media sites support ephemeral sharing and augmenting photos and videos with stickers (e.g., Instagram and Facebook Stories), Snapchat was the only widely-used platform with an API that supported importing external stickers for use within the platform. Consequently, participant recruitment became challenging, as eligible participants needed to both be active enough on Snapchat and have enough experience collecting personal data to minimize novelty effects of sharing data ephemerally. Beyond limiting eligible participants, the nature of social media ecologies~\cite{zhao2016social} meant that data was absent or difficult to integrate into conversations participants had with their Snapchat contacts off the platform.

Aligning with the expectation of polish, research suggests the need for tools that share personal data to support customization of visuals, text, and numbers to avoid repetition in sharing~\cite{epstein2020yarn, liu2017supporting, munson2012exploring}. To support this goal, the authors implemented a relatively large number of stickers (39) and customization options (e.g., sticker color, unit of measurement like ``steps'' and ``miles''). Other customization options, such as font and text color, required development work that fell out of scope (e.g., wrapping text within a sticker, ensuring legibility). Participants appreciated the sticker and customization options and used them widely. However, multiple participants still feared that the content they shared would eventually become boring if they continued using SnapPI beyond the study duration, and suggested that additional stickers and customization options would help.

Accessing participants' use of SnapPI required violating both personal and platform privacy expectations. Because stickers were exported from SnapPI to Snapchat, SnapPI's telemetry was unable to record how participants embedded the sticker in their Snap, whether participants sent a Snap with the sticker at all, who they sent it to, or how audience members responded. The authors, therefore, asked participants to send their Snaps to an account associated with the research team and took screenshots for later analysis. Although this enabled participants to filter out Snaps containing photos or information they did not want to disclose to the researchers, some participants frequently forgot to share their Snaps with the research team. Additionally, the ephemerality of Snapchat creates expectations that posts will disappear~\cite{xu2016automatic}. The researchers taking screenshots for later analysis violated this expectation, and may have influenced how participants perceived or used the platform.

\subsection{Expressive Biosignals}
\begin{figure}[t]
    \footnotesize
      \centering
          \includegraphics[width=0.98\linewidth]{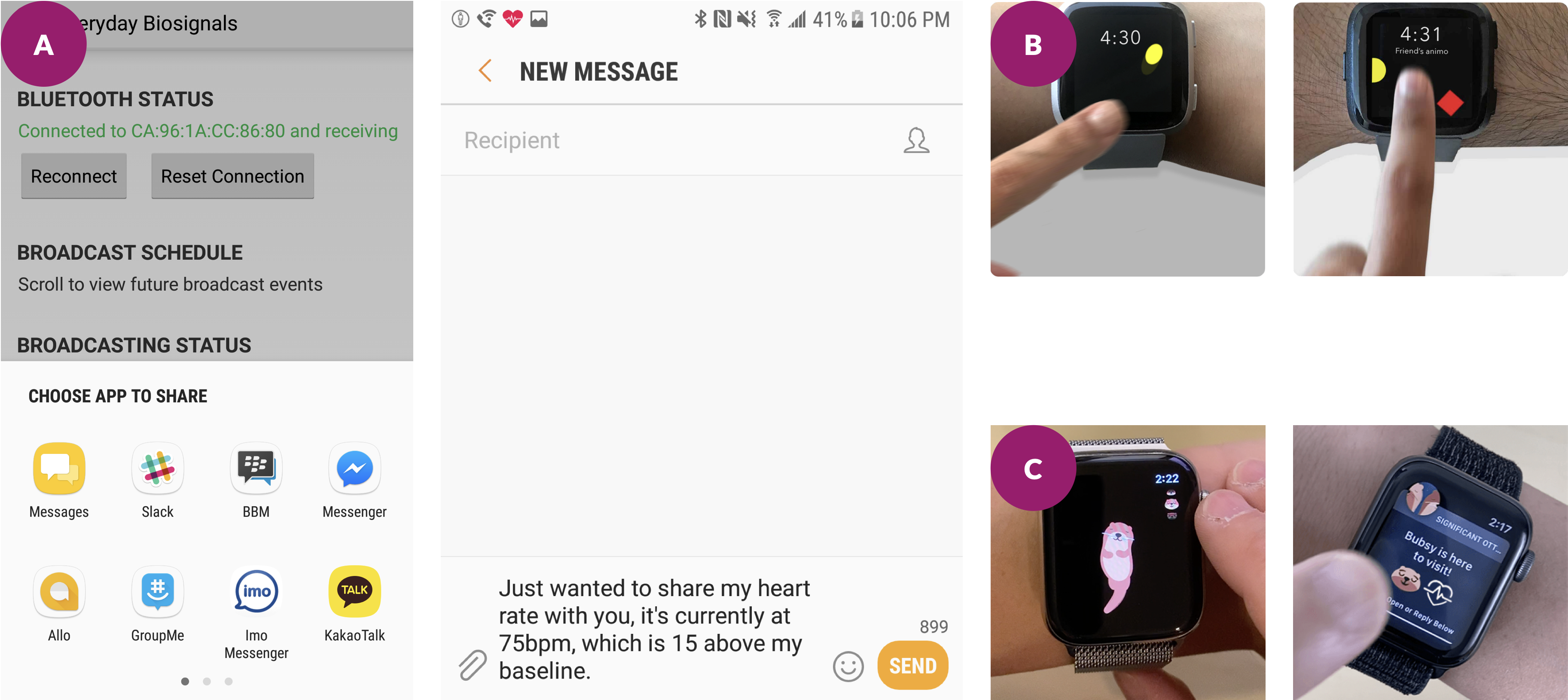} 
    \caption{The (A) Everyday Biosignals system supported people in sharing their heart rate via messaging apps, enabling understanding of biosignal use in everyday conversation within chosen communication channels. Conversely, (B) Animo and (C) Significant Otter implemented standalone systems for communicating biosignals, giving greater design flexibility by removing the format constraints of existing channels.}
        \label{fig:biosignals}
\end{figure}

Our second case study focuses on prototypes that integrate \textit{expressive biosignals}, or the display of sensed physiological data in social interactions~\cite{liu2020fostering}. Motivated by the limited nonverbal cues available in remote communication, as well as the increasing ubiquity of consumer-grade wearable sensors, this line of research introduced biosignals as a new social cue that technology can afford. 
Liu et al. investigated several research questions around understanding people's decisions behind sharing such data and their social impact in dyadic communication. For the purposes of this paper, we describe and compare three expressive biosignals systems they built that vary in their degree of piggyback prototyping: an ``Everyday Biosignals'' text messaging system~\cite{liu2017supporting}, the Animo Fitbit Versa watchface~\cite{liu2019animo}, and the Significant Otter iPhone/Apple Watch app~\cite{liu2021significant} (Figure~\ref{fig:biosignals}).

\subsubsection{Using piggyback prototyping}
Everyday Biosignals is an Android app that prompts users to text their heart rate to their contacts using a messaging app of their choice (e.g., built-in SMS, WhatsApp, Snapchat). Liu et al. built and deployed this system to explore people's sharing motivations and behaviors around biosignals like heart rate~\cite{liu2017supporting}. The app connects to the Mio Alpha 2 watch via Bluetooth LE in order to continuously read heart rate, which is used to send notifications (i.e., when heart rate is high or low compared to the user's baseline) asking the user if they would like to share their current heart rate with a contact. If the user selects ``yes,'' they can then select a messaging app of their choice, which the system opens and pre-populates with editable text containing their heart rate in beats per minute (bpm). As such, the system piggybacks off of these existing messaging apps to enable users to message their heart rate to others. The authors found that with this system, people shared their heart rate in order to express their emotions and give status updates, but the heart rate bpm was not engaging by itself and required people to apply their own meanings to it.

The authors' decision to piggyback off of existing messaging apps was due to the ability of users ``\textit{to share with their contacts in a naturalistic fashion using their typical communication channels.}'' Since people use different communication apps for different people or contexts~\cite{nouwens2017whatsapp}, leveraging their existing apps reduced the need for participants to learn and incorporate a new channel into their communication, and allowed them to more naturally express themselves on platforms they typically used with any of their contacts. This enabled high ecological validity, where participants considered heart rate sharing in light of their real-world communication practices. Subsequently, like SnapPI, users embedded their heart rate into conversations they already have with others. The researchers could thus address one of their key questions: ``\textit{how can people meaningfully express their heart rate to others?}''

Because Everyday Biosignals piggybacked off of other communication apps,  Liu et al. asked participants to send the researchers screenshots of the conversations they had in their communication apps when they shared their heart rate. The researchers could therefore observe how people spoke about their heart rate with others, including how they edited the pre-populated heart rate text, provided context or explanation for the text, or conversed about the data. 
At the same time, using screenshots limited the ability to instrument how participants used the various chat apps to converse around their biosignals. For instance, the screenshots were sometimes insufficient to understand the context of the conversations that participants had around the shared biosignals.

\subsubsection{Building a standalone system}
In contrast to Everyday Biosignals, Liu et al. also created two new systems which did not piggyback off of existing social computing systems: Animo~\cite{liu2019animo} and Significant Otter~\cite{liu2021significant}. 
Both are smartwatch apps that enable users to share their heart rate with each other in the form of an animated avatar. The authors found that sharing heart rate in these ways enabled people to feel connected with close others in a lightweight way throughout their daily lives. Unlike Everyday Biosignals, these systems are new social computing apps that the authors built and deployed to understand the effects of biosignals sharing.

Piggyback prototyping with Everyday Biosignals had the benefits of exploring biosignals as integrated into natural texting behaviors. In contrast, Animo and Significant Otter brought the opportunity to explore biosignals sharing as its own communication channel. In particular, developing standalone applications enabled greater design freedom, without being tied to the format of existing messaging apps (i.e., text). Since Everyday Biosignals' raw text heart rate in bpm was not very engaging, the authors designed more playful representations for biosignals – specifically, shapes or otters that animated according to high or low heart rate. Additionally, exploring these representations outside of back-and-forth text conversations allowed biosignals to act as lightweight cues on their own,
which helped participants keep up-to-date with each other and start new conversations at later times. The authors also explored the potential for sharing biosignals on a new form factor: smartwatches. This form factor enhanced ease-of-use, as participants could quickly glance at and tap their avatar on their watchface to send it, as well as feelings of presence, where its location on the wrist gave a sense of being physically ``touched.'' Leveraging this form factor would not have been possible through piggybacking, as there were very few existing social watch apps at the time, and none that allowed for customizing the shared watchface via piggybacking. Finally, developing a new app created a unique channel between participants and their partner, which enabled a more intimate dyadic communication experience.

At the same time, qualitative results from these works reveal some drawbacks to building a standalone app. A few participants wished that the app was more integrated with their existing messaging apps, such that they could use their established forms of communication with their partner or other people outside of the study. For example, they described wanting to paste their avatar into their texts rather than only view and send it on their watch. Some participants emulated this by sending their avatar and immediately sending a follow-up text to explain that avatar. This is similar to Everyday Biosignals, where people used the conversation to provide context for ambiguous biosignals. Thus, while building standalone systems had many benefits that outweighed these drawbacks, Liu et al. still highlight integration with existing social computing systems as a major design implication.

\subsection{Mindful Garden: Co-located Social Interaction in Augmented Reality}

\begin{figure}[t]
    \footnotesize
      \centering
          \includegraphics[width=0.98\linewidth]{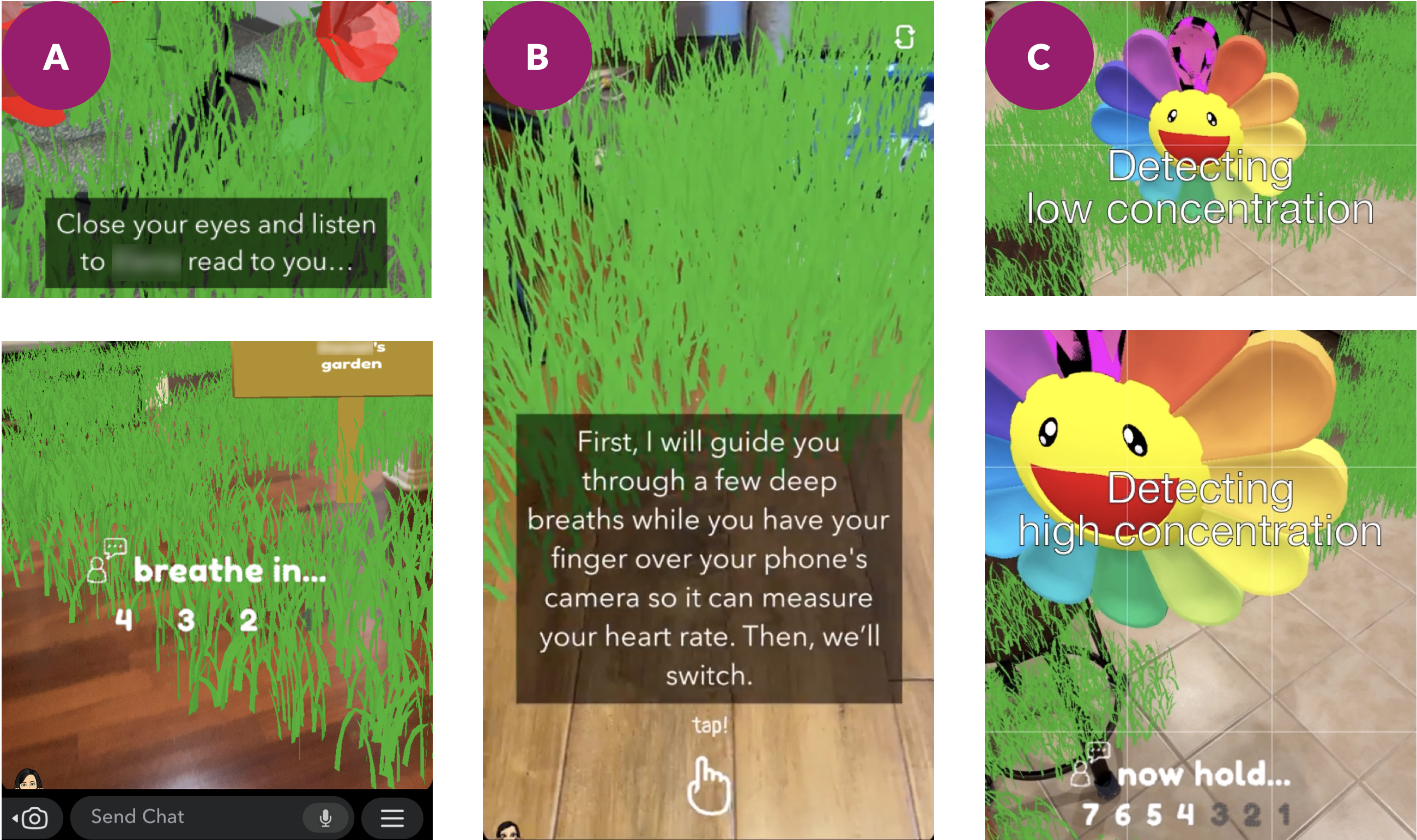} 
    \caption{Mindful Garden piggybacks off of Snapchat, leveraging the platform's Augmented Reality framework to (A) support a co-located mindfulness session where a virtual garden is populated by flowers correlating with biosignals detected from the camera. (B) In the first version, heart rate is detected by having the person being guided place their finger over the phone's camera. (C) In the second version, more sophisticated biosignals are detected from an external device, requiring data to be transferred to Snapchat through a covert channel.}
        \label{fig:mindfulgarden}
\end{figure}

Mindful Garden is an ongoing project exploring how augmented reality on mobile phones and headsets, in conjunction with biosignal data, can help people who are co-located in a physical space connect and share with one another~\cite{mindfulgarden}. Co-located mindfulness meditation has the potential to more deeply connect the people meditating, giving space for them to jointly influence their environment. Biosignals can offer feedback on how mindful a person is being, and can help make people feel closer to one another. In Mindful Garden, pairs of people guide each other through a breathing exercise, lasting about 1 minute per person. In augmented reality, both participants are presented a garden containing flowers which grow correlated with each participant's biosignals (Figure~\ref{fig:mindfulgarden}). Mindful Garden is built as a Lens on Snapchat with Lens Studio~\cite{snaplensstudio} to leverage the platform's support for co-located augmented reality and minimize the setup burden of installing a new app.

Two versions of Mindful Garden are currently being developed. In the first version, the number of flowers are correlated with the heart rate of the person receiving the meditation guidance. To infer heart rate, person receiving meditation guidance places their finger over their phone's camera for the system to detect differences in average pixel color. This allows the system to operate entirely on a mobile device within the Snapchat app. In the second version, participants' flowers are correlated with heart rate and brain wave data from a Muse 2 headset, with indicators like the flower's growth rate, color, and number of petals correlated with specific brain waves.

The benefits of piggybacking Mindful Garden off of Snapchat have been primarily technical. Preliminary findings suggest that piggybacking was helpful for easing the burden of setup, as the system was distributed as an add-on to the social platform (via a Snap Code) rather than requiring separate installation. It also allowed for building off of the well-tested and complex development environment within Snapchat for augmented reality, reducing the technical burden of making the Lenses.

The challenges of piggyback prototyping in Mindful Garden have also been primarily technical. As of the time of writing, Lens Studio does not support HTTP requests for accessing external data, beyond information Snap already collects about a person (e.g., name, birthday) or their environment (e.g., altitude, temperature, time, weather)~\cite{snapdynamicinformation}. To incorporate biosignals, both versions of Mindful Garden therefore make use of input signals supported by Lens Studio. The first version leverages raw pixel data from the camera to directly measure heart rate. For the second version, the researchers tested a few techniques to transfer biosignal data from the Muse 2 headset to the lens. One technique involved displaying a digital tag (e.g., a QR code, a recognizable object) that encoded the data on a device such as a tablet in the co-located environment. While this technique was technically feasible, it required participants to closely point the camera at the tag to read it. This interfered with the ability to render a garden in the environment, as most of the visible environment was the tag, and also would distract the participant from the mindful activity. The technique ultimately used was to have the device in the environment play a tone at an inaudible frequency which encodes the biosignal values from the Muse device, which the Lens can interpret.

These technical challenges have resulted in less desirable user experiences. In the first version, leveraging the camera as an input source resulted in relatively low accuracy of heart rate measurement. Consequently, people using the app tend to interpret the flower growth as random, rather than being connected to their heart rate. While heart rate measurement has been more accurate in the second version, the indirect transfer of the biosignal has only allowed  a few discrete values to be transmitted. Because audio detection in the Lens was primarily intended to capture audible noise, a relatively narrow band of audio frequency was both inaudible and detectable by the Lens. This led to limitations around the fidelity of data that could be transmitted, only allowing biosignals such as level of concentration to be regarded as either ``low'' or ``high'', rather than the continuous input that the Muse 2 headset provides. This discretization has constrained some ideas around encoding the biosignals into the garden design.

\section{Discussion}

Through reviewing the goals, benefits, and tensions of piggyback prototyping social communication experiences, we have extended Grevet \& Gilbert's focus on assessing critical mass of interest~\cite{grevet2015piggyback} to demonstrate how the method can promote other research goals including the ordering and filtering of social content or formation of affinity groups. We articulate how the method can increase the ecological validity of research studies by leveraging familiarity, enabling more interesting and realistic social relationships, and structuring participant data collection. We also describe challenges to deploying the method, including technical limitations, expectations and disclosure concerns. Here, we briefly outline points of consideration when deciding whether to piggyback off of an existing social system. We then advocate for wider use of the method, and finally, describe a few potential research agendas to support piggyback prototyping.

\subsection{Considerations When Deciding Whether to Piggyback}

We suggest three considerations when deciding whether piggybacking off of an existing social system is more effective for answering a research question than developing a social system from scratch. 
\subsubsection{\textcolor{black}{Is my research question better answered by leveraging an existing platform or starting from scratch?}}

\textcolor{black}{A first consideration is whether piggyback prototyping is beneficial in answering the intended research question. We believe that piggyback prototyping has significant advantages over starting from scratch for answering research questions around how people perceive a new technique or capability. Specifically, it can allow for a deeper understanding of the everyday considerations, constraints, and circumstances people will inevitably face if the technique being prototyped were present in social computing systems people use regularly. However, piggybacking can constrain the design space of social interactions to existing tools' paradigms, limiting piggybacking if the research question proposes a radically different approach to communication. For example, it might prove difficult or limiting to design for sharing some types of content through piggybacking given the capabilities of existing systems, like prototyping social experiences in virtual reality.}

\textcolor{black}{A related question to consider is whether people are already communicating on an existing platform, or are interested in doing so.} The main benefit of piggyback prototyping is minimizing the gap between people's everyday use and research use, suggesting value in selecting platforms where the relevant stakeholders have already congregated or are interested in congregating. 
Further, considering whether people's use of the platform aligns with the research goal can help reduce friction to use, as people have preferred platforms for communicating with specific people or about specific topics~\cite{nouwens2017whatsapp}. For example, a researcher interested in testing new mechanisms for sharing messages of affection may benefit from piggybacking off of a direct communication tool like SMS rather than a productivity communication tool like Slack, while the reverse might be preferable for testing a creative ideation communication workflow.

A potentially valuable practice is to survey or interview the target audience about platform interest and appropriateness before designing, which can influence the choice of platform to piggyback off of~\cite{lambton2019whatfutures, lukoff2018tablechat}. For conducting studies, researchers should also consider whether they will be able to recruit participants who are interested in and able to communicate on the platform and meet other inclusion criteria. Piggybacking off of multiple platforms may allow for greater flexibility, but comes with an additional development burden.

\subsubsection{Do any platforms allow for and simplify the design approach I am considering?}

With benefits assumed, a key practical question is whether it is even possible to extend an existing system as desired. In particular, we recommend establishing whether a given customization is fundamentally possible before investing design and development resources into implementing the method. Many systems leverage hacks, loopholes, or other workarounds to accomplish the goal, such as transmitting  biosignals over audio in Mindful Garden. In other systems, a less-optimal user experience was supported because the intended customization was not possible~\cite{zhang2018making}. In all of these cases, an important consideration is to what extent the workaround will be noticeable by a potential participant or user, or to what extent it will negatively impact the experience, as limited or impacted participant use can impact researcher's ability to answer their research questions.

A range of technical options can be used for piggyback prototyping. Researchers prototyping on top of systems on the web might consider browser extensions. Browser extensions are well-suited for piggybacking as they are not dependent on the underlying system having extensible features or APIs and only impact the person who has the extension installed, allowing for asymmetric experiences. In extreme circumstances, researchers could leverage pixel-based techniques to enable advanced customizations~\cite{dixon2010prefab}. \textcolor{black}{In these cases, it is worth considering whether piggybacking reduces the development burden beyond building a social computing system from scratch.} Alternatively, for contributions where empirical evaluations are not needed, it may be sufficient to provide a technical sketch of how such an extension would work on a particular platform~\cite{zhang2020policykit}.

Prior uses of piggyback prototyping also emphasize that many research questions can be answered solely through repurposing a platform's built-in mechanics, such as creating a researcher-organized Facebook Group~\cite{macleod2016asynchronous, maestre2018defining} or structuring a Google Doc to support the desired communication task~\cite{o2018suddenly}. If existing mechanics are sufficient for achieving the intended research goal, it can help reduce development burden and increase participant familiarity.

\textcolor{black}{Beyond viability, it is important to consider whether piggybacking is worth the risks inherited from building on an existing platform. A piggyback prototype study can become completely upended if a commercial API is adjusted to revoke a necessary feature, if a website is adjusted in a way that breaks a browser extension, or if a system is taken down entirely. While deployments of standalone social computing systems in the field are frequently disrupted by researcher-created bugs~\cite{siek2014field}, the researcher is not as beholden to the whims of the creators of the system being piggybacked on, and has enough control to address issues mid-deployment. On the flip side, piggybacking can inherit some reliability from well-tested commercial systems, such as minimizing concerns about whether messages will be delivered from one person to another.}

\textcolor{black}{As a generalization, our inclination is that features of a system are likely safe to piggyback on if non-researchers rely on the techniques. For example, SnapPI leveraged Snap's Creative Kit, which advertisers use to create custom stickers and lenses for their websites and products~\cite{snapcreativekit}. This instills  trust that, in theory, some warning would be issued prior to sunsetting the API. For platforms that may frequently change or A/B test their interfaces or features with minimal or no warning (e.g., websites used by browser extensions), we believe that the
relatively fast development cycles of most piggyback prototypes (e.g., a year or two) and relatively short durations of deployment studies (e.g., weeks or a month) can lower risks. However, we expect that the risks are likely less worth inheriting in social computing systems which aim to exist long-term and undergo frequent iteration. For example, reflecting on a decade of Python Tutor~\cite{guo2013online, guo2015codeopticon, guo2020learnersourcing}, Guo identified that the reliance on extremely minimal software dependencies enabled greater reliability of the social system~\cite{guo2021ten}.}

\subsubsection{Does piggybacking in this domain introduce undue privacy or ethical concerns?}

As discussed, the nature of piggybacking on social platforms provides access to the conversations that participants are having, which in many cases is part of the research data. If those conversations are sensitive, it may not be desirable to disclose them. Additionally, as  pointed out by prior work, people's public or semi-public behaviors may become part of the research data~\cite{munson2015effects, grevet2015piggyback}. We recommend that researchers creating piggyback prototypes clarify what information about engagement is being collected, both with participants and with anyone who may engage with participant-generated content. In cases where the privacy or ethical risks of piggybacking exceed the benefit of leveraging an existing platform, a standalone system may need to be developed or the research question may need to be refocused.

\subsection{Advocating for piggybacking and piggybackable platforms}

In outlining a set of benefits that piggyback prototyping provides for answering a range of CSCW systems questions, we argue that the community could benefit from broader use of the method relative to designing single-purpose research systems. We also suggest that making more social systems piggybackable can enable further research and experimentation, and deepen our knowledge of how to support people's social communication. We advocate for supporting extensibility of both commercial social communication systems and research prototypes.

For commercial systems, supporting extensibility can allow the broader community to envision, design, evaluate, and advocate for communication techniques that are not on the critical path of the platform. For example, the Boomerang extension for schedule-sending emails inspired a technique for improving how people use Gmail to communicate with one another, which was eventually integrated into the platform~\cite{boomerang}. Similarly, TweetDeck's dashboard was eventually acquired by Twitter, and some features were subsequently integrated into the native app~\cite{tweetdeck}.

For research systems, the benefits roughly align with the benefits of supporting extensibility or comparison of other research systems. For example, if a research app that implements a new kind of communication were released as open-source or with extensible APIs, it could support more direct comparisons, such as allowing A/B testing between the previous and new technique.

In spite of the benefits, it is important to acknowledge the risks and downsides of supporting this sort of extensibility. For research systems, requiring extensibility further increases the bar of contributing to a social system. The support and maintenance of such a system, while admirable and valuable, may not align well with the expectations of the early-career academics who frequently build them~\cite{guo2021ten}. For commercial systems, supporting extensibility inevitably creates an opportunity for others to misuse or abuse the platform, with Cambridge Analytica being one prominent example~\cite{cambridgeanalytica}.

\subsubsection{Acknowledging the difficulty of piggybacking social systems}

Although we see the benefit in piggybacking as a method for evaluating new concepts in social systems, we also wish to point out the difficulties of executing it. While researchers may be able to set up some expectation that issues will arise in field deployments of prototypes~\cite{siek2014field}, participants using piggyback prototypes are occasionally less forgiving of issues with the prototype. These issues also have the potential to impact participants' social relationships as part of everyday use. They may worry how their friends and family might interpret their use- or non-use when things go wrong, such as when messages are sent multiple times or not sent at all. Rather than simply inconveniencing the participant and impacting the researcher's ability to collect useful perspectives, piggyback prototypes embed the researcher in the participants' social system with all of its complexities and expectations. Because piggybacked systems often leverage workarounds or hacks to implement their novel social communication techniques, these technical and usability issues may be more frequent than in other kinds of prototypes. Addressing these challenges can require additional testing or development beyond the high expectations of deployment studies, or may not be addressable at all.

It may be possible to partially couch or focus participant's feedback through more targeted approaches than strictly measuring or experimentally comparing use with or without the prototype. Qualitative methods like interviews or surveys can help give participants space to articulate what felt unpolished or could be streamlined alongside feedback on the core idea that the prototype aimed to evaluate.

Relatedly, Bernstein et al. note difficulties articulating the novelty of a social computing systems~\cite{bernstein2011trouble}. We echo Bernstein et al.'s suggestion that piggyback prototypes, like other social computing systems, can be evaluated on their \textit{social contributions}, considering whether the prototype enables some new social interaction. Past that point, as Bernstein et al. suggest, any voluntary use by participants can be viewed as a success, and should be viewed by the research community as such.

\subsection{New Research Agendas Suggested By Piggyback Prototyping}

The widespread interest and use of piggyback prototyping points to valuable new research agendas and directions for future CSCW systems researchers, specifically in the need for richer toolkits. 
One potentially valuable direction for future research is in tools to support researchers and developers looking to extend existing platforms. These tools would primarily aim to lower the threshold to developing piggyback prototypes~\cite{myers2000past}, providing APIs which implement common features needed for piggybacking or contribute methods for piggybacking without programming. \textcolor{black}{While prior piggyback prototypes occasionally leveraged existing third-party toolkits, such as existing chatbot frameworks ~\cite{zhang2018making, lukoff2018tablechat}, we found that most implemented their prototype either from scratch (e.g., via manipulating a web page's DOM) or with the help of the platform's official APIs. Other spaces have demonstrated the utility of researcher-created toolkits for common problems, such as researchers piggybacking off of TurKit~\cite{little2010turkit} to prototype crowdsourcing workflows for Amazon Mechanical Turk~\cite{bernstein2010soylent, bigham2010vizwiz, chilton2013cascade}.}

Because many piggyback prototypes aim to support creation of shareable content, research contributing tools for authoring content on social systems could support future piggyback prototypes. For example, many researchers have suggested that their prototype would benefit from more customizable shareable content ~\cite{munson2012exploring, epstein2020yarn, griggio2021mediating}. Toolkits could help both researchers and participants by providing common customization options such as editable text, image layout, colors, emoji, and fonts, and support exporting them to social platforms. However, understanding the design requirements for such a tool would require further research. Additionally, tools could aggregate authoring features from different commercial APIs to allow studies to recruit participants who have different platform preferences. In doing so, tool designers could explore ways of reducing the burden to developers answering whether platforms allow for the planned customization, such as considering how to surface the content authoring limitations of different platforms.

When the platforms being piggybacked on do not provide APIs for accessing or manipulating shared content, piggyback prototypes often develop pipelines for accessing and editing content such as browser extensions. We observed some patterns implemented across a few different prototypes, such as accessing the video of a livestream~\cite{lu2018streamwiki, yang2020snapstream}. By understanding the needs of developers looking to extend these  systems and developing tools for supporting them, researchers could contribute paradigms for extending existing platforms.

Tools can also help researchers configure existing social communication systems around affinity groups, supporting paradigms like the ARC method~\cite{macleod2016asynchronous, maestre2018defining, prabhakar2017investigating} and WhatFutures~\cite{lambton2019whatfutures, lambton2021blending}. Tools could promote some of the effective design choices made in prior instantiations of the methods, such as bringing up how the names of groups can impact privacy~\cite{maestre2018defining}. Tools could also support automating tasks that APIs allow for, but prior researchers have manually undertaken to avoid development burden. For example, tools could provide mechanisms for scheduling daily activities in the ARC method rather than requiring researchers to post them~\cite{macleod2016asynchronous}, and tools could automatically collect what participants communicate to one another in the group's social feed.

\textcolor{black}{Beyond toolkits, it is also worth considering whether methods like Wizard of Oz could be used to avoid technical barriers to piggyback prototyping while still evaluating the core research questions~\cite{gould1985designing}. For example, a novel approach for summarizing content might be able to avoid a limitation around programmatic access to existing posts in a social computing system by a researcher manually copying posts over, within reason. Similarly, studying a new kind of social content could avoid limitations around programmatic authoring of content, by giving participants text, images, etc. which can be copied directly into the platform. Designing such Wizard of Oz methods, and demonstrating scenarios where they could support rapid ideation and evaluation of design ideas for social computing systems, is a promising direction for complementing piggyback prototypes.}

\section{Conclusion}

In reconsidering and expanding the definition of piggyback prototyping, we demonstrate that in addition to assessing critical mass of interest on a topic in a social network, the method is helpful for answering research questions on the design of social communication features. We describe how the method provides ecological benefits over developing standalone prototypes, but also introduces ethical challenges around who can access participant conversations during studies. We encourage social computing systems researchers to consider whether the method is better-suited to answering their research questions than developing a standalone system, and suggest the research agenda of developing tools to reduce the burden of applying the method.

\begin{acks}
We thank our coauthors of SnapPI, Expressive Biosignals, and Mindful Garden and participants at the Social Computing Systems Summer Camp for initial discussions on this topic. This work was supported in part by the National Science Foundation under award IIS-1850389, the UCI Council on Research, Computing, and Libraries (CORCL), and the Snap Creative Challenge on the Future of Co-located Social Augmented Reality.
\end{acks}

\bibliographystyle{ACM-Reference-Format}
\bibliography{x-references}


\end{document}